\documentstyle[12pt]{article}

\catcode`\@=11
\long\def\@makefntext#1{
\protect\noindent \hbox to 3.2pt {\hskip-.9pt
$^{{\ninerm\@thefnmark}}$\hfil}#1\hfill}                

 \def\@makefnmark{\hbox to 0pt{$^{\@thefnmark}$\hss}}  

\def\ps@myheadings{\let\@mkboth\@gobbletwo
\def\@oddhead{\hbox{}
\rightmark\hfil\ninerm\thepage}
\def\@oddfoot{}\def\@evenhead{\ninerm\thepage\hfil
\leftmark\hbox{}}\def\@evenfoot{}
\def\sectionmark##1{}\def\subsectionmark##1{}}

\newcounter{sectionc}\newcounter{subsectionc}\newcounter{subsubsectionc}
\renewcommand{\section}[1] {\vspace{0.6cm}\addtocounter{sectionc}{1}
\setcounter{subsectionc}{0}\setcounter{subsubsectionc}{0}\noindent
	{\bf\thesectionc. #1}\par\vspace{0.4cm}}
\renewcommand{\subsection}[1] {\vspace{0.6cm}\addtocounter{subsectionc}{1}
	\setcounter{subsubsectionc}{0}\noindent
	{\it\thesectionc.\thesubsectionc. #1}\par\vspace{0.4cm}}
\renewcommand{\subsubsection}[1]
{\vspace{0.6cm}\addtocounter{subsubsectionc}{1}
	\noindent {\rm\thesectionc.\thesubsectionc.\thesubsubsectionc.
	#1}\par\vspace{0.4cm}}

\newcounter{appendixc}
\newcounter{subappendixc}[appendixc]
\newcounter{subsubappendixc}[subappendixc]

\renewcommand{\appendix}[1] {\vspace{0.6cm}
	\refstepcounter{appendixc}
	\setcounter{figure}{0}
	\setcounter{table}{0}
	\setcounter{equation}{0}
	\renewcommand{\thefigure}{\Alph{appendixc}.\arabic{figure}}
	\renewcommand{\thetable}{\Alph{appendixc}.\arabic{table}}
	\renewcommand{\theappendixc}{\Alph{appendixc}}
	\renewcommand{\theequation}{\Alph{appendixc}.\arabic{equation}}
	\noindent{\bf Appendix \theappendixc #1}\par\vspace{0.4cm}}

\def\abstracts#1{{
	\centering{\begin{minipage}{30pc}\tenrm\baselineskip=12pt\noindent
	\centerline{\tenrm ABSTRACT}\vspace{0.3cm}
	\parindent=0pt #1
	\end{minipage}}\par}}

\renewenvironment{thebibliography}[1]
	{\begin{list}{\arabic{enumi}.}
	{\usecounter{enumi}\setlength{\parsep}{0pt}
\setlength{\leftmargin 1.25cm}{\rightmargin 0pt}
	 \setlength{\itemsep}{0pt} \settowidth
	{\labelwidth}{#1.}\sloppy}}{\end{list}}

\newcounter{itemlistc}
\newcounter{romanlistc}
\newcounter{alphlistc}
\newcounter{arabiclistc}

\newcommand{\fcaption}[1]{
	\refstepcounter{figure}
	\setbox\@tempboxa = \hbox{\tenrm Fig.~\thefigure. #1}
	\ifdim \wd\@tempboxa > 6in
	   {\begin{center}
	\parbox{6in}{\tenrm\baselineskip=12pt Fig.~\thefigure. #1}
	    \end{center}}
	\else
	     {\begin{center}
	     {\tenrm Fig.~\thefigure. #1}
	      \end{center}}
	\fi}

\newcommand{\tcaption}[1]{
	\refstepcounter{table}
	\setbox\@tempboxa = \hbox{\tenrm Table~\thetable. #1}
	\ifdim \wd\@tempboxa > 6in
	   {\begin{center}
	\parbox{6in}{\tenrm\baselineskip=12pt Table~\thetable. #1}
	    \end{center}}
	\else
	     {\begin{center}
	     {\tenrm Table~\thetable. #1}
	      \end{center}}
	\fi}

\def\@citex[#1]#2{\if@filesw\immediate\write\@auxout
	{\string\citation{#2}}\fi
\def\@citea{}\@cite{\@for\@citeb:=#2\do
	{\@citea\def\@citea{,}\@ifundefined
	{b@\@citeb}{{\bf ?}\@warning
	{Citation `\@citeb' on page \thepage \space undefined}}
	{\csname b@\@citeb\endcsname}}}{#1}}

\newif\if@cghi
\def\cite{\@cghitrue\@ifnextchar [{\@tempswatrue
	\@citex}{\@tempswafalse\@citex[]}}
\def\citelow{\@cghifalse\@ifnextchar [{\@tempswatrue
	\@citex}{\@tempswafalse\@citex[]}}
\def\@cite#1#2{{$\null^{#1}$\if@tempswa\typeout
	{IJCGA warning: optional citation argument
	ignored: `#2'} \fi}}

\def\fnt#1#2{\footnotetext{\kern-.3em
	{$^{\mbox{\sevenrm #1}}$}{#2}}}

 1
 1
 1

\font\tenbf=cmbx10
\font\tenrm=cmr10
\font\tenit=cmti10

\font\ninerm=cmr9

\begin{document}

\rightline{LANCS-TH/9516 (1995), e-Print Archive: hep-ph/9506336}

\centerline{\tenbf IS ELECTROWEAK BARYOGENESIS
CLASSICAL? \footnote{\ninerm This talk was given at the 7th Adriatic
Meeting on Particle Physics: Perspectives in Particle Physics '94
13-20 Sep 1994,  Islands of Brijuni, Croatia}}
\vspace{0.8cm}
\centerline{\tenrm TOMISLAV PROKOPEC
\footnote{\ninerm Based on collaboration with
Michael Joyce and Neil Turok}
\footnote{\ninerm e-mail: T.Prokopec@lancaster.ac.uk}}
\baselineskip=13pt
\centerline{\tenit Lancaster University, Physics Department}
\baselineskip=12pt
\centerline{\tenit Lancaster, LA1 4YB, England, UK}
\vspace{0.3cm}
\vspace{0.9cm}
\abstracts{
In this lecture first I present a brief genesis of the ideas on the
electroweak baryogenesis and then I
focus on a mechanism in which
the source of $CP$ violation
is a $CP$-violating field condensate which could  occur, for example,
in multi-higgs extensions of the Standard Model. In the limit of a thick bubble
wall one finds a classical
force acting on particles proportional to the mass squared and the $CP$
violating phase. One can study this
force in the fluid  approximation in which the effects of transport
and particle decays can be taken into account.
A novelty in this talk is
generalization of the problem to the relativistic velocity.
There is a regime in which
the final formula for the baryon asymmetry has a rather simple form.
}

\vfil
\rm\baselineskip=14pt

\section{Prelude ..}

In this lecture I will present
some of the recent developments in  understanding
baryogenesis at the electroweak phase transition.

I will focus on the work
I have done in collaboration with Michael Joyce and Neil Turok.
The main idea is rather simple:

\vspace{0.4cm}
\noindent
{\it In the
semiclassical approximation baryogenesis can be described by a
$CP$ violating classical field condensate. The effect of this condensate can be
studied in the fluid approximation which  takes account of both
transport and particle decays.}
\vspace{0.2cm}

\indent

I will also present some of the hystorical  ideas  that `seeded' our work as
well some of the recent developments that I find related or interesting.
(I cannot hope for a lack of personal bias in my choice.)

\section{Sakharov's Conditions and the Electroweak Theory}

The baryon-to-entropy ratio of the Universe one would like to explain  is
\begin{equation}
{n_B\over s}= (4-7) 10^{-11}
\label{eq:baryon to entropy ratio}
\end{equation}
This number is a
consistency constraint for nucleosynthesis calculations  be in agreement
with the  observed primordial elements' abundances. For a pedagogical review
see \cite{Steigman}, for recent developments \cite{Krauss}.

`Before Sakharov' baryogenesis was a fine tuning problem of the intial
conditions of the Universe. Sakharov realized \cite{Sakharov} that
baryons can be generated {\it dynamically\/}
 provided the following conditions are satisfied

$\bullet\triangleright$ {\it Condition 1:\/}
there exist processes that violate baryon number

$\bullet\triangleright$
{\it Condition 2:\/} system is out of thermal equilibrium

$\bullet\triangleright$
{\it Condition 3:\/} there exist both $C$ and $CP$ violating processes

The necessity of {\it Condition\/} 1 is obvious.

An elegant proof of {\it Condition\/ 2} can be found in
\cite{DimopoulosSusskind}. In thermal equilibrium the hamiltonian $\cal H$
 is $CPT$ invariant, and the baryon number $B$ is odd under $CPT$
\begin{equation}
(CPT) \; {\cal H}\; (CPT)^{-1} ={\cal H}\,,\qquad
(CPT) \; B\; (CPT)^{-1} = - B
\label{eq:hamiltonian CPT invariance}
\end{equation}
In thermal bath baryon number is given by a thermal average
\begin{equation}
\langle B\rangle = {\rm Tr} \left [ {\rm e}^{-\beta {\cal H}} B
\right ]
\label{eq:baryon number thermal average}
\end{equation}
We can now insert $(CPT) (CPT)^{-1}=1$ into the trace and use
Eq.~(\ref{eq:hamiltonian CPT invariance}) to get
\begin{eqnarray}
\langle B\rangle & = & {\rm Tr} \left [ (CPT) {\rm e}^{-\beta {\cal H}} B
(CPT)^{-1}\right ]\nonumber\\
& = & {\rm Tr} \left [  {\rm e}^{-\beta {\cal H}} (CPT) B
(CPT)^{-1}\right ] = -\langle B\rangle
\label{eq:baryon number CPT consideration}
\end{eqnarray}
so $\langle B\rangle =0$.

Next I present an argument which indicates the necessity of $CP$ violation, as
stated by {\it Condition 3}.
(A similar proof as above applies, but I want to avoid
the use of thermal equilibrium.)
Recall since the net baryon number is defined as the difference between the
number of baryons and antibaryons $B=b-\bar b$ and under $CP$:
$b\rightarrow \bar b$ and $\bar b\rightarrow b$, we have
\begin{equation}
(CP) \; B\; (CP)^{-1} =- B
\label{eq:baryon number under CP}
\end{equation}
In case there is no $CP$ violation the processes which create net $b$ occur
with the same rate as the processes that create net $\bar b$ and {\it no\/}
net $B$ will be created.

{\it Are the Sakharov conditions realized in the Standard Model? }

We start with the baryon number violation.
The Weinberg-Salam theory has on the quantum level
a remarkable axial current  anomaly
through
which baryon number current is violated
\begin{eqnarray}
\partial_\mu J_B^\mu & = &
N_F{g_w^2\over 32\pi^2} [-{W^a}_{\mu\nu} \tilde {W}^{a\,\mu\nu} ]
+N_F {g_1^2\over 32\pi^2} B_{\mu\nu} \tilde {B}^{\mu\nu}\nonumber\\
& \equiv & N_F {g_w^2\over 32\pi^2}\left [
- \partial_\mu K^\mu +\partial_\mu k^\mu \right ]
\nonumber\\
K^\mu & = & {1\over 2}\epsilon^{\mu\nu\rho\sigma} {\rm Tr}\left [
W_\nu\partial_\rho W_\sigma +{2\over 3} i g_w W_\nu W_\rho W_\sigma
\right ]   \nonumber\\
k^\mu & = & {1\over 2} \epsilon^{\mu\nu\rho\sigma}\left [
B_\nu\partial_\rho B_\sigma \right ]
\label{eq:anomaly for baryons}
\end{eqnarray}
where $N_F=3$ is the number of families,
$W_{\mu \nu}$ and $B_{\mu \nu}$ and $W_{\mu}$ and $B_{\mu}$
denote the $SU(2)_L$ and $U(1)_Y$  field strengths and fields, respectively,
and $\tilde {W}^{\mu\nu}={1\over 2}\epsilon^{\mu\nu\rho\sigma} {W}^{\mu\nu}$
denotes the dual field strenght.
The fact that $\partial_\mu K^\mu$ and $\partial_\mu k^\mu$
are total derivatives does not mean that this
anomalous  current violation can be gauged away without any physical
consequences by merely shifting the current.
The resolution is in the nontrivial
topological structure of the theory.
The integral of
Eq.~(\ref{eq:anomaly for baryons}) defines how are
the topology changes in the gauge fields related to the changes of
the baryon number:
\begin{equation}
\Delta B = N_F\Delta n_{CS}\,,\qquad
n_{CS}= - {g_w^2\over 32\pi^2} \int K^\mu  dS_\mu
\label{eq:topological change}
\end{equation}
At zero temperature quantum tunnelling can generate baryons \cite{tHooft}
but the rate is tiny $\sim e^{-4\pi/\alpha_w}$ --  it would not create even
one baryon in all of the (visible) Universe!

The discovery of the sphaleron \cite{Manton}, a classical solution to the
dimensionally reduced action which carries a half topological charge and as
such, it was argued, may mediate baryon number violating processes, lead to a
belief that the finite temperature baryon number violating rate may be
unsuppressed \cite{KuzminRS} which would make the electroweak
baryogenesis possible. This was in fact proven by Arnold and McLerran in
\cite{ArnoldMcLerran} where they showed that in the unbroken phase the
number of sphaleron transitions (per unit volume and time) reads
\begin{equation}
{\Gamma_{sph}\over V}\sim {E_{sph}^7\over T^3} e^{-{E_{sph}\over T}}\,,\qquad
E_{sph}= A(\lambda/g_w^2) {2 m_w (T) \over \alpha_w}
\label{eq:sphaleron rate broken phase}
\end{equation}
where $A(\lambda/g_w^2)\sim 1.5 - 2.7$; in fact for physically
 interesting range of $\lambda/g_w^2$ to a good approximation
$A\sim 1.7$. I have not bothered to quote all pre-exponential factors
in this relationship; they can be found in \cite{ArnoldMcLerran}.
The main point is the sphaleron rate is exponentially suppressed by the
Boltzmann factor with the sphaleron free energy in the exponent, which
is defined as the saddle point  of the dimensionally reduced three dimensional
action.

An important condition which ought to be satisfied in order for any electroweak
baryogenesis model be viable is

$\bullet\triangleright$ {\it Sphaleron erasure:} The baryons generated at the
transition must not be washed out by the subsequent sphaleron transitions in
the broken phase \cite{ShaposhnikovSphaleron}.

Roughly speaking this means that the sphaleron rate per unit time
$\Gamma_{sph}$ in the broken
phase ought to be smaller than the expansion rate of the Universe $H$.
More precisely, since at the time of the phase transition completion,
the higgs expectation value $\phi_0=\langle \phi\rangle$ still grows quite
rapidly on the expansion time scale, sphalerons are active only
a small  portion of the expansion time after the completion \cite{GuyMoore}.
With this in mind
one gets slightly less stringent bound on the sphaleron energy
\begin{equation}
E_{sph}\geq 35 T
\label{eq:sphaleron erasure}
\end{equation}
Taking account of $M_w= g_w \phi_0/2$ and the dependence of the higgs
expectation value on the couplings (which is roughly
$\phi_0\sim g^3 T/\lambda_T$) one gets an upper bound on the
higgs mass. It turns out that  with the two loop effetive potential this
leads to a bound: $m_H<30GeV$. If one allows for nonperturbative effects this
bound becomes less stringent. Recall that the current constraint on the
higgs mass
is $m_H>60 GeV$ for the Standard Model higgs. Fortunately this does not mean
the end of electroweak baryogenesis. Since we are interested in an extended
higgs sector which contains $CP$ violation
and the constraints on the mass of the lightest higgs are much less stringent,
and the above calculation should be altered accordingly,
I believe that at the moment no serious discrepancy exists.

An important calculation of the `sphaleron' rate in the unbroken phase is still
missing due to the ill understanding of the infrared sector of gauge
theories. Indeed it is believed that magnetic fields,  which are
involved in the processes, are screened at the scale  of the infamous
magnetic mass: $m_{mag}\sim g_w^2 T$, which specifies the `magnetic'  field
correlation length  $\xi\sim 1/g_w^2 T$.

We now present a simple estimate of the rate.
There will be instantons of all sizes $\rho\in \{0, \xi\}$ which represent
tunneling trajectories in the configuration space with the barrier height of
order $1/\alpha_w \rho$. The smallest barrier is for large instantons
$\rho\sim \xi$ with the energy  $\sim T$ so that there is no exponential
suppression to tunnelling; the rate is then determined by the prefactor which
is  of order
$\xi^4$, hence $\Gamma_{sph}/V\sim (\alpha_w T)^4$.

A number of numerical studies
\cite{GregorievAmbjorn} supports this naive argument
\begin{equation}
{\Gamma_{sph}\over V}= \kappa_{sph} (\alpha_w T)^4 \,,\qquad
\kappa_{sph}\sim 0.1 - 1
\label{eq:sphaleron rate symmetric phase}
\end{equation}
There are however problems with the simulation. Ambjorn {\it et al.\/} use the
real time classical theory which is plagued with ultraviolet divergences;
the authors hope that the lattice {\it cut-off\/} takes care of that. Also the
Gauss contraint is not naturally imposed. I will take
Eq.~(\ref{eq:sphaleron rate symmetric phase}) as a guidance to the true rate.

Since the symmetric phase `sphaleron' rate
Eq.~(\ref{eq:sphaleron rate symmetric phase}) is much faster
than the broken phase rate
Eq.~(\ref{eq:sphaleron rate broken phase}) an efficient baryogenesis
mechanism should generate baryons in the symmetric phase.
This is indeed possible if one takes account of particle transport.

Regarding {\it Condition 2\/} perturbative calculations
\cite{ArnoldEspinosaFodor} valid for a rather light
higgs particle $m_H\leq 70$\mbox{GeV} indicate that the phase transition is
first order and likely proceed {\it via\/} bubble nucleation providing the
required departure from thermal equilibrium. To resolve the problem fully one
requires to solve for nonperturbative effects which is possible only
by using  lattice simulation. Recently \cite{FarakosShaposhnikov}
it has been shown that the dimensionally reduced
theory is well suited for studying equilibrium properties of the transition
in a lattice simulation. A preliminary
result \cite{Shaposhnikovprivate} indicates that the
phase transition becomes second order for  the higgs mass above about
$80\mbox{\rm GeV}$.

What about {\it Condition 3\/}?
$C$ is maximally violated. For example no right handed neutrino has been
observed.

The main problem for the  Standard Model baryogenesis is a small $CP$
violation. A natural measure is the $CP$ violating phase from the
Kobayashi-Maskawa matrix $\delta _{CP}\sim 10^{-19}$, and since it is usually
argued that $n_B\propto \delta_{CP}$ it seems without an additional CP
violation the Standard Model baryogenesis is out! This however may not be true.
Recently a couple of attemts have been made to resolve this formidable problem.
Shaposhnikov and Farrar \cite{ShaposhnikovFarrar} (inspired by a modest $CP$
violation of the neutral kaon) have realized that under
certain conditions the thermal  reflection off the bubble wall may lead to
a net baryonic current due to the difference in the tree level light quarks'
masses. This model, since in its original formulation
takes no account of the
possible loss of quantum coherence in the reflection off the wall,
has been recently jeopardized
\cite{Gavela}, \cite{HuetSather}.
Another notable attempt to solve the problem of a small $CP$ violation is by
Nasser and Turok \cite{NasserTurok}. Based on a linear response analysis
they argue that there might be instability
to formation of a $Z$ field condensate on the bubble wall caused by the
chiral charge deposit from the reflected top quarks. This
condensate as we will see below may  generate a net baryon number.
The longitudinal $Z$ field domains of
the opposite sign form indiscriminately on the wall
and no net baryon number is formed  unless there is a nonzero $CP$ violation.
This gives an
advantage to the formation of a particular sign condensate. Diffusion of
the one sign domains against the other enhances the bare
$CP$ violation. If this mechanism works (as it may for a very heavy top quark)
 it would result in $n_B\propto \sqrt \delta_{CP}$.

{\it How can one bias baryon number production?}

If there is for example a term in the free energy (possibly dynamically
generated) that couples to the baryon current
\begin{equation}
\Delta F = a_\mu  J_B^\mu
\label{eq:term that biases baryon production}
\end{equation}
then a classical formula of the near equilibrium statistical physics gives
\begin{equation}
\dot {n}_B= -\Gamma_{sph}\mu_B\,,\qquad \mu_B= {\delta F\over \delta n_B}=a_0
\label{eq:formula of statistical physics}
\end{equation}
where $\mu_B$ is  the chemical potential for the baryon number.
(In this light {\it Condition 2} of Sakharov asserts that in thermal
equilibrium $\mu_B=0$.)
Indeed baryon number production is biased if  on average $a_0\neq 0$.
This reasoning
sets the stage for
the next section.

\section{The ideas of Cohen, Kaplan and Nelson}

There are two ideas of relevance for my work:

$\bullet\triangleright$ {\it spontaneous\/} baryogenesis and

$\bullet\triangleright$ {\it charge transport}

The idea of the {\it spontaneous} baryogenesis
in its intial form \cite{CohenKaplan}
is rather simple. Cohen and Kaplan assumed that at some early stage in the
Universe $CPT$ symmetry of the hamiltonian was temporarily violated by a
term of form:
\begin{equation}
{1\over M} \partial_\mu\theta J_B^\mu \rightarrow
{1\over M} \dot \theta  (n_b-n_{\bar b})
\,,\qquad M\geq 10^{13} \mbox {\rm GeV} \,,\quad T < M
\label{eq:spontaneous bg CPT violation}
\end{equation}
The implication $(\rightarrow)$ applies for a homogeneous field $\theta$.
The {\it ilion} field
$\dot \theta$ (the name is supposed to bear reminiscence of the axion) acts
as a chemical potential for baryon number which, when the expansion of
the Universe is
taken ito account, drives $B\neq 0$, hence the name {\it spontaneous}.

The idea got a new impetus when Dine {\it et al.\/} \cite{DineHSS}
realized that  a similar term arises on one-loop level in theories which
contain a singlet field ({\it e.g.\/} supersymmetric theories).
The effective action then acquires a term
\begin{equation}
{1\over 3 M_{susy}} \partial_\mu s J_B^\mu
\label{eq:spontaneous bg singlet field}
\end{equation}
which may lead to a {\it spontaneous} baryogensis at a first order
electroweak phase transition. Note that in this scenario the electroweak
transition is used to drive the system out of equilibrium.
A similar term occurs at one
loop in two higgs doublet theories with $CP$ violation in the higgs
sector \cite{TurokZadrozny}. But then
Cohen, Kaplan and Nelson
\cite{CKNspontaneous} realized that in these theories even at
the {\it tree\/} level there is a term
\begin{equation}
 y_F \partial_\mu \theta J_F^\mu \,,\qquad
J_F^\mu = \bar\Psi\gamma^\mu \gamma_5 \Psi
\label{eq:spontaneous bg hyperchage}
\end{equation}
which couples the fermionic hypercharge
current $J_F^\mu$, which is closely related to the baryonic current,
with a $CP$ violating  field $\theta$, the gauge invariant
relative higgs phase. $\dot\theta$
acts as a chemical potential for the baryon number. On an expanding  bubble
wall $\dot \theta$ is predominantly of one sign, as given by the
{\it classical\/} equations of motion, and hence {\it it\/} drives the
sphaleron  rate in a definite direction
creating baryons. As originally designed this mechanism is not manifestly mass
suppressed, {\it i.e.\/} baryon number does not vanish in the limit
when fermion masses go to {\it zero.\/} In this limit the chiral symmetry is
restored and baryon creation {\it must\/}  stop. I will present a resolution
\cite{JPTconstraints} of this paradox below.

The idea  of {\it charge transport\/} was designed
\cite{CKNtransportOLD} in a model with
heavy (majorana) neutrinos and explicit $CP$ violation in the coupling
constants. In this model $CP$ violation takes place on the bubble wall. As a
result of a $CP$ violating reflection on the wall an
axial current is injected and {\it transported} into the symmetric phase
where the sphaleron rate is unsuppressed
Eq.~(\ref{eq:sphaleron rate symmetric phase}).
This mechanism takes advantage of
a large $CP$ violation on the bubble wall and of a fast baryon number violating
rate in the symmetric phase so it is potentially very efficient. Indeed in the
subsequent work Cohen, Kaplan and Nelson \cite{CKNtransport}
have applied the idea to the top quark
and managed to generate baryon asymmetry as large as $\sim 10^{-6}\theta_{CP}$.
Since the injected reflected chiral top current is the result of a coherent
quantum reflection, it is essential that top quarks do not scatter
on the bubble wall. Also a significant reflection occurs only when the wall
is thinner than $m_t^{-1}$. Both of these lead to a rather stringent constraint
on the wall thickness $L_w\leq 2/T$ which would require a very strong first
order phase transition and this is,
according to perturbative calculations of
the effective thermal higgs potential \cite{ArnoldEspinosaFodor},
highly unrealististic.

We have realized \cite{JPTleptons} that this idea might have a `brighter
future' if aplied to the particles that couple weakly
to the wall and scatter infrequently. Our favorite
candidate is the right-handed  $\tau$-lepton since it has a long scattering
lenght and diffuses very efficiently. We have shown that with
$\theta_{CP}\sim 1$ the baryon  number produced may be consistent with
Eq.~(\ref{eq:baryon to entropy ratio}).

\section{Two higgs doublet model and $Z$ condensate}

In order to study the effects of $CP$ violating condensate
on the particles' dynamics on the wall I will rewrite the
Lagrangian in a convenient form so that the interesting $CP$ violation is
expressed in the form of a $CP$ violating $Z$ field condensate.

Consider first the higgs kinetic term. I can pick
a gauge in which the background classical solution reads
\begin{equation}
\Phi_1={1\over \sqrt{2} } \left ( {0\atop \phi_1 e^{i\theta_1}} \right )
\,,\qquad
\Phi_2={1\over \sqrt{2} } \left ( {0\atop \phi_2 e^{i\theta_2}} \right )
\label{eq:higgs expectation values}
\end{equation}
In what follows I will ignore the electromagnetic fields ($U(1)_{EM}$
remains unbroken so there is no obvious reason for a condensate to appear),
the charged $W^{\pm}$ gauge bosons and the higgs excitations both neutral and
charged. Then the higgs kinetic  term can be diagonalized
\begin{eqnarray}
 |{\cal D}\Phi_1|^2 +|{\cal D} \Phi_2|^2 & = &
 {1\over 2} {\phi_1^2 \phi_2^2 \over \phi^2}
\left [\partial_\mu(\theta_1 -\theta_2)\right ]^2
+ {\phi^2\over 2}   \left [
{g\over 2}  Z_\mu -{1\over \phi^2}
\left (\phi_1^2 \partial_\mu \theta_1 + \phi_2^2 \partial_\mu \theta_2 \right )
\right ]^2 \nonumber\\
&\equiv & {\phi_1^2 \phi_2^2 \over 2 \phi^2}
\partial_\mu \theta \partial^\mu \theta +
{\phi^2 \over 2} \left ({g\over 2}  Z_{GI}\right ) ^2 \nonumber\\
g Z_\mu & = & g_2 W_\mu^3 - g_1 B_\mu\,,\qquad
\phi^2 = \phi_1^2+\phi_2^2\,,\quad  g^2=g_1^2+g_2^2
\label{eq:higgs kinetic term}
\end{eqnarray}
where I ignored the kinetic terms for $\phi_1$ and $\phi_2$.
Note that when either of the higgs {\it vevs\/} vanish the gauge invariant
relative phase  $\theta$ loses its meaning, as it should.
Both $\theta$ and $Z_\mu^{GI}$ are gauge invariant, and we take them
as our definition of the $CP$ violating condensate fields.

Consider now the fermionic part of the Lagrangian.
Assume for simplicity that only $\Phi_1$ couples to the fermions
{\it via\/} the Yukawa terms. The phase $\theta_1$ can be removed by performing
an anomaly free rotation proportional to $[T^3-Y+\frac{1}{2}(B-L)]$  on the
fermions (and also on the Higgs field) at the cost of introducing a new term
in the fermionic kinetic term
\begin{equation}
\bar\Psi i \gamma^\mu
\left ( \partial_\mu - ig_A \tilde{Z}_\mu\gamma_5 \right )\Psi
- m\bar\Psi\Psi
\label{eq:fermionic terms}
\end{equation}
where $g_A= +g/4$ ($g^2=g_1^2+g_2^2$) for up-type quarks and (left-handed)
neutrinos, and $g_A = -g/4$ for down-type quarks and charged leptons, and
\begin{equation}
g_A \tilde{Z}_\mu= g_A Z_\mu^{GI} - {\phi_2^2 \over 2 \phi^2}
\partial_\mu\theta
\label{eq:gauge invariant Z condensate}
\end{equation}
The advantage of this form is that I have reduced the effect
the higgs phase to a pure gauge $Z$ field condensate  in the presence of
a (real) mass $m=y\phi_1/\sqrt 2$. Since the axial symmetry is broken
this field produces an interesting physical effect which we want to study.
This formulation is suitable for studying the effect of a $CP$ violating
condensate in the semiclassical limit.

An alternative formulation is to study the classical evolution of the $CP$
violating phase accross the wall which actually neglects the plasma effects.
The two-higgs-doublet higgs potential
may contain $CP$ violating terms
\begin{equation}
\lambda_5 \left (
{\cal Re}\, [\Phi_1^\dagger \Phi_2 ] - v_1 v_2 \cos \xi_0
\right )^2
+ \lambda_6 \left (
{\cal Im}\, [\Phi_1^\dagger \Phi_2 ] - v_1 v_2 \sin \xi_0
\right )^2
\label{eq:two higgs doublet potential}
\end{equation}
so that the classical evolution of the relative higgs phase
\begin{equation}
\partial_\mu \partial^\mu \theta -\phi^2 {v_1 v_2\over \phi_1 \phi_2}
\left [
\lambda_5 \cos\xi_0 \sin\theta +\lambda_6\sin\xi_0 \cos\theta
+ {\phi_1 \phi_2 \over v_1 v_2} (\lambda_6-\lambda_5)\sin\theta \cos\theta
\right ] =0
\label{eq:equation of motion for theta}
\end{equation}
For a slow moving wall $\partial_\mu \partial^\mu \theta\sim 0$ can be
neglected and we have for the symmetric and broken phases
\begin{eqnarray}
\tan\theta  =  -{\lambda_6\over \lambda_5}\tan\xi_0\,, & &\qquad
\phi_1=\phi_2=0
\nonumber\\
\cos\xi_0\tan\theta +{\lambda_6\over \lambda_5 }  \sin\xi_0
+ \left ({\lambda_6\over \lambda_5}-1 \right ) =0\,, & &
\phi_1 =  v_1\,,\quad \phi_2=v_2
\label{eq:equations theta symmetric and broken}
\end{eqnarray}
so that only if $\lambda_6 \neq \lambda _5$ there is a net change in
$\theta$ accross the wall, {\it i.e.\/} $CP$ violation and $Z$ condensate.
For a small $CP$ violation
\begin{equation}
\Delta \theta \simeq -{(\lambda_6-\lambda_5)  \lambda_6\cos\xi_0 \over
 \lambda_6^2\cos^2\xi_0 + \lambda_5^2  \sin^2 \xi_0}
\label{eq:delta theta small CP violation}
\end{equation}
which is an upper limit to change in $\theta$ given by
Eq.~(\ref{eq:equation of motion for theta}).
How this may drive baryogenesis was first studied in \cite{TurokZadrozny}
and then in  \cite{CKNspontaneous}.

\section{A free particle on the bubble wall}

We are now ready to study the motion of a free particle on a bubble wall with a
$Z$ field condensate, {\it i.e.\/} $CP$ violation.
Considering a planar wall stationary so that
$\tilde Z^\mu= (Z^0,0,0,Z_3(z))$.

First using Eq.~(\ref{eq:fermionic terms})  we write the Dirac equation  (in
the Fourier space)
\begin{equation}
\left [
P\!\!\!\!\slash - g_A \tilde {Z}\!\!\!\!\slash\gamma_5 - m
\right ]\Psi=0
\label{eq:dirac equation with Z field}
\end{equation}
which in the chiral representation
(where $\Psi\sim [\Psi_R, \Psi_L ]$)
can be broken into two equations
for the two component right-handed $\Psi_R$ and left-handed $\Psi_L$ spinors
\begin{eqnarray}
\left [ E- g_A Z^0-
\vec\sigma\cdot (\vec P- g_A \vec Z )\right ]\Psi_R +m\Psi_L & =0 &
\nonumber\\
m\Psi_R+ \left [ E+ g_A Z^0+
\vec\sigma\cdot (\vec P+ g_A \vec Z )\right ]\Psi_L & = & 0
\label{eq:dirac equation for two spinors}
\end{eqnarray}
where  we used $\gamma_5\Psi_{R,L}=\pm \Psi_{R,L}$. Setting the determinant
to zero leads to the dispersion relation which is simple in the wall frame
where $\tilde Z^\mu= (0,0,0,Z(z))$
\begin{eqnarray}
E = \left [p_\perp^2+
\left (\sqrt{p_z^2+m^2} \mp g_A Z\right )^2\right ]^{1/2}
 \qquad S^z=\pm {1\over 2}
\label{eq:dispersion relation}
\end{eqnarray}
$S_z$ is the component of the spin in the $z$ direction,
measured in the frame in which $p_\perp$ vanishes.
In the WKB approximation, this dispersion relation
accurately describes particles as they move  across
a bubble wall - the local eigenstates in
Eq.~(\ref{eq:dispersion relation}) shall form the basis of our
treatment.
The particles we are most interested in for baryogenesis are
left handed particles ({\it e.g.\/} $t_L$'s),
and their antiparticles ($\overline{t_L}$'s, which are
right-handed), because these couple to the chiral anomaly.
For large $|p_z|$, these are easily identifiable in terms
of the eigenstates in
Eq.~(\ref{eq:dispersion relation}).
Note that they couple {\it oppositely} to the $Z$ field.

The group velocity of a WKB wavepacket is determined from the dispersion
relation by
$v_z = \dot z = \partial E / \partial p_z$, and
energy conservation $\dot E =0$ implies
$\dot p_z = -\partial_z E$.  From these it
is straightforward to calculate the acceleration
\begin{equation}
{d v_z \over dt} = -{1\over 2} {(m^2)'\over E^2} \pm{(g_A Z m^2)'
\over E^2 \sqrt{E^2-p_\perp^2} }  +o(Z)
\label{eq:acceleration}
\end{equation}
where $E$ and $p_\perp$ are constants of motion.

This chiral force provided by the $Z$ field
effectively produces a potential well which draws
an excess chiral charge onto the wall, and leads to a compensating
deficit in a  `diffusion tail' in front of the wall.
There is net baryon production because $B$ violation
is suppressed on the bubble wall.

\section{Fluid equations}

I now seek to describe the particle excitations with
dispersion relations
Eq.~(\ref{eq:dispersion relation}) as
classical fluids.
I focus on
particles with large  $|p_z|\sim T>>m$ for the following reasons:
they dominate phase space,  the
$WKB$ approximation is valid, and
the dispersion relation simplifies so one can
identify approximate chiral eigenstates.
The $S_z=+{1\over2}$, $p_z<0$
branch, and the $S_z=-{1\over 2}$, $p_z>0$ branch
constitute one, approximately left
handed fluid $L$, and the other two branches an approximately  right-handed
fluid $R$.

The Boltzmann equation is:
\begin{equation}
d_t f \equiv \partial_t f + \dot z \partial_z f + \dot p_z
\partial_{p_z} f = - C(f)
\label{eq:boltzmann equation}
\end{equation}
where $\dot z =\partial_{p_z} E$ and $\dot p_z= -\partial_z E$
are calculated from the
Hamilton equations and $C(f)$
is the collision integral.

In order to study a system close to equilibrium, fluid approximation is
reasonably accurate, provided

$\bullet\triangleright$ the scatterings within a fluid are more frequent then
between any two fluids,

$\bullet\triangleright$ the distortions
 in the momentum space are small so that
the momentum expansion is well behaved (see below); this is the case
{\it e.g.\/} for a thick bubble wall.

In the plasma frame the fluid {\it Ansatz\/} for the distribution function is
\begin{equation}
f={1\over {{\rm e}^{E+\delta }\pm 1}}
\,,\qquad \delta = -\mu +p^\nu \delta_\nu + ...
\,,\qquad \delta^\nu =(-\delta T, 0,0,v)
\label{eq:distribution function}
\end{equation}
where I have for convenience set $T=1$ and truncated the expansion at the
vector term. Since the dispersion relationship
Eq.~(\ref{eq:dispersion relation})
is obtained in the wall frame we transform energy and momentum
\begin{equation}
E\rightarrow \gamma_w (E+v_w p_z)
\,,\qquad p_z\rightarrow \gamma_w (p_z+v_w E)
\,,\qquad p_\perp\rightarrow p_\perp
\label{eq:lorentz transform E p}
\end{equation}
which gives Eq.~(\ref{eq:distribution function}) in the wall frame
\begin{equation}
f={1\over {{\rm e}^{\gamma_w(E+v_w p_z) +\delta }\pm 1}}
\,,\qquad \delta = -\mu -\gamma_w (E+v_w p_z) \delta T -
\gamma_w(p_z+v_w E) v
\label{eq:distribution function wall frame}
\end{equation}
There is a subtlety in deriving the fluid equations.
In the zero mass limit there must be no observable effect (particle
number perturbation) due to a nonzero $Z$ field since it is pure gauge.
However $d_t p_z =-\partial_z E=-(p_z\pm g_A Z)/E) \partial _z (g_A Z)$ would
generate a chemical potential. This is what Cohen, Kaplan  and Nelson
indentified as the term that drives spontaneous baryogenesis.
Dine and Thomas \cite{DineThomas} and also Joyce, Turok and
myself \cite{JPTconstraints} have pointed out that there must be a
mass squared suppression.
Therefore this term is a pure gauge artefact. The proper (gauge covariant)
form of $f$   includes the kinetic momentum and a shifted `physical'
chemical potential
%
\begin{equation}
k_z=p_z\pm g_A Z \,,\qquad
\hat \mu = \mu \pm \gamma_w v_w g_A Z
\label{eq:kinetic momentum and physical chemical potential}
\end{equation}
where the signs are chosen appropriately.
The source term
\begin{equation}
d_t \gamma_w(E+v_w k_z)= \gamma_w v_w \dot k_z
=\gamma_w v_w (\partial_z E\pm g_A \dot Z)
\label{eq:source term}
\end{equation}
vanishes in the zero mass limit as it should.
I can now write down the linearized form of the $l.h.s.$ of
Eq.~(\ref{eq:boltzmann equation}) as
\begin{equation}
d_t f=f' (\gamma_w (E+v_w p_z)) \left [
\gamma_w v_w \dot k_z
-d_t \mu -\gamma_w (E+v_w p_z) d_t \delta T -
\gamma_w(p_z+v_w E) d_t v
\right ]
\label{eq:distribution function wall frame ii}
\end{equation}
with $d_t \rightarrow \partial_{p_z} E \partial _z
\simeq {p_z\over E} \partial _z $ in the wall frame.

Rather then studying the full fluid equations I will simplify the problem
 by neglecting the temperature fluctuation. This is reasonably accurate
when the scatterings that destroy $\delta T$ are
much more efficient than the decay processes that destroy $\mu$.
In the electroweak plasma this is satisfied  to a
good approximation.
For a detailed treatment of the complete fluid equations  see \cite{JPTthick}.

In order to obtain the fluid equations I integrate
Eq.~(\ref{eq:distribution function wall frame ii}) in the wall frame over
${\int d\!\!\!^-}^3p$  and $\int {d\!\!\!^-}^3 p \gamma_w (p_z+ v_w E) $.
I have chosen this  Lorentz-like combination of
$E$ and $p_z$ since in this case
the collision term has a simple form when evaluated in the  fluid frame:
only the velocity perturbation is damped and both
chemical potential and temperature perturbations drop out.
After a lenghty algebra
I arrive at the fluid equations  for particle minus
antiparticle perturbations ($\mu =\mu(L)-\mu(\,\overline{L}\,)$,
$v =v(L) -v(\,\overline{L}\,)$)
in the rest frame of the wall
\begin{eqnarray}
-a_{23} v_w\gamma_w \hat\mu^\prime
+ \gamma_w  \frac {1} {3} v^\prime
 & = &
-{\Gamma}_{\mu 1} \left [ \hat \mu \right ]
-\Gamma_{\mu 1}^* \left [ \hat \mu +2 \gamma_w v_w g_A Z \right ]
\nonumber\\
\label{eq:fluid equation i} \\
a_{34}\gamma_w \hat\mu^\prime -\gamma_w v_w v'
+ F_3
& = & -\Gamma_v v
\nonumber\\
F_3=3 a_{14} v_w (g_A Z m^2)^\prime  & &
\!\!\!\!\!\!\!\!\!\!\left (1-{v_w\over 2}\ln{1+v_w\over 1-v_w}\right )
\label{eq:fluid equation ii}
\end{eqnarray}
where $[\hat \mu]$ denotes the signed sum of
chemical potentials for particles participating in the reaction,
prime denotes $\partial_z$,
$a_n=(n! \zeta _n/2\pi^2) [1-1/2^{n-1}]$ ($n>1$),
$a_1=(\ln (2/m)/2\pi^2) [\ln m]$,
$a_{23}=a_2/a_3=(\zeta_2/3\zeta_3)[2/3]$,
$a_{34}=a_3/a_4=(\zeta_3/4\zeta_4) [6/7]$,
$a_{14}=a_1/a_4=(\ln 2/24)[8/7]$
 and $\zeta$  is the Riemann
$\zeta$-function; to obtain the constants $a_{ij}$
for bosons one should simply drop the  factors in square brackets.

There are two sources of baryogenesis in these equations

$\bullet\triangleright$ the classical force term
$\propto  (g_A Z m^2)^\prime  $ in the second equation and

$\bullet\triangleright$ the hypercharge violating processes
$\Gamma_{\mu 1}^* \left [2 \gamma_w v_w g_A Z \right ]$

The rates ${\Gamma}_{\mu 1}$ and  ${\Gamma}_{v}$  damp the chemical potential
and velocity perturbations respectively. If one for example considers the top
quark baryogenesis, the main tree level procesess that contribute to
${\Gamma}_{\mu 1}$ are the  helicity flipping top-gluon scattering
with the emission of a charged higgs particle or a charged $W$ boson and
permutations of these.
The process with the higgs emission (absorbtion) is faster due to its large
Yukawa coupling. For quarks in addition there is also the anomalous strong
sphaleron process that flips chirality.
The hypercharge violating helicity flipping processes are $(m/\pi T)^2$
suppressed and therefore rather slow (see \cite{JPTthick}).
In this letter I shall focus on studying the effect of the classical force.
In order to simplify the discussion I shall set all decay rates to zero.
This is justified provided during the wall passage a small portion of the
seeded perturbation decays {\it i.e.\/}
\begin{equation}
{L\over v_w}<< \Gamma_{\mu 1}, \Gamma_{\mu 1}^*
\label{eq:no decay condition}
\end{equation}
Also it is important that  decays do not cut-off the diffusion tail in front of
the wall. I will comment more on it below.

I now set all decay terms to zero
$\Gamma_{\mu 1}, \Gamma_{\mu 1}^* \rightarrow 0 $.
Eq.~(\ref{eq:fluid equation i}) can be integrated to give
\begin{equation}
  v = 3 a_{23} v_w\hat\mu
\label{eq:integral of fluid equation i}
\end{equation}
since at $+\infty $ both $\hat \mu= v=0$.
Eq.~(\ref{eq:fluid equation ii}) can then be written as
\begin{eqnarray}
\bar D\hat\mu^\prime + v_w \hat \mu & = & s'\equiv
 - {F_3 \over 3 a_{23} \Gamma_v}
\nonumber\\
\bar D= D \left (1-3 v_w^2{a_{23}\over a_{34}}\right )\,, & &
 \qquad D={a_{34}\over 3 a_{23}\Gamma_v}
\label{eq:fluid equation for mu}
\end{eqnarray}
where $D$ is the (plasma frame) diffusion constant. The inclusion of the
temperature perturbations changes $D$ only slightly; in that case
$D=1/3\Gamma_v$.
We have calculated \cite{JPTthick} the contribution of the quark-quark
gluon exchange tree-level  diagram and found $D\sim 6/T$, a similar
contribution is expected from the quark-gluon scattering diagram, which would
half the diffusion constant \cite{MooreProkopec}.

Note that close to the speed of sound $v_w\sim v_s= 1/\sqrt 3$
in this approximation the fluid becomes  {\it stiff\/},
{\it i.e.\/} transport is suppressed ($\bar D/v_w<< L$) and
there is only local baryogenesis:
\begin{equation}
\hat\mu= {s'\over v_w}=-{1\over 3 a_{23} v_w}{F_3\over \Gamma_v}
\label{eq:mu at speed of sound}
\end{equation}
which is suppressed as $\bar D/Lv_w$.
This result should be taken with caution since the limit
$\bar D\sim 0$ might signify the
breakdown of the fluid approximation.
In a more careful treatment one should include other types of perturbations,
most notably $\delta T$.
If one does so one finds that the perturbation at the source may be
transported back to distances $\sim 1/\Gamma_v$.
So generically there is no diffusion
in front of the  wall, but particles {\it do\/} diffuse behind the wall.

In order to write  a general solution to Eq.~(\ref{eq:fluid equation for mu})
we write down the solution using the Greens function method
\begin{eqnarray}
\partial_x G+{v_w\over \bar D} G=\delta (x-y)
\,,\quad G =\theta (x-y)\exp \left [ - {v_w\over \bar D} (x-y) \right ]
\,,\qquad \hat \mu = \int G s'
\label{eq:solution for mu using greens function}
\end{eqnarray}
for $\bar D>0$.
This form of the solution is quite useful since using partial integration
it can be expanded in powers of $\bar D/ v_w L$ for thick walls, or
 $L v_w/\bar D $ for thin walls.
I leave this as an exercise. The Greens function in
Eq.~(\ref{eq:solution for mu using greens function})
is the  response to a $\delta$-function source and
as such illustrates transport properties of the full solution
Eq.~(\ref{eq:solution for mu using greens function}):
in this approximation when no
particles decay all of the sourced perturbation gets transported
into the diffusion tail in
front. This tail extends to $\bar D/v_w$ in front of the source which is
large in the limit of a small velocity. A particle spends on average particle
\begin{equation}
\tau_{diff}\sim {\bar D\over v_w^2}
\label{eq:diffusing time}
\end{equation}
diffusing in front of the wall. If this time is larger than any od the
(zero mass) decay rates, then the diffusion tail will be suppressed
as $\Gamma_\mu \bar D/ v_w^2$
and accordingly the source for baryogenesis will be reduced.

In order to solve for the baryon number I may integrate
Eq.~(\ref{eq:formula of statistical physics}) and find that it suffices to know
the integral of $\hat\mu$  in front of the wall (for non-local baryogenesis) or
on the wall (for local baryogenesis). Integrating
Eq.~(\ref{eq:fluid equation for mu}) and assuming
efficient transport such that $v_w L/\bar D<< 1$,
I find that {\it on the wall}
,
\begin{equation}
\hat\mu = {s\over \bar D} =
-\frac{2 \ln 2}{3\zeta_3} \gamma_w v_w g_A Z m^2 V(v_w)
\qquad \gamma_w V(v_w)= {1-{v_w\over 2}\ln{1+v_w\over 1-v_w}
\over  1-3 v_w^2{a_{23}/a_{34}} }
\label{eq:shifted mu}
\end{equation}
In order to obtain the baryon number I as anticipated integrate
Eq.~(\ref{eq:formula of statistical physics}) with all appropriate constants
to find in the plasma frame
\begin{equation}
-{1\over v_w\gamma_w} \int\dot n_B=n_B =  {3\over 2} N_C \int
{\Gamma_{sph}\over V}
(\mu_{t_L} - \mu_{\overline{t}_L})
\label{eq:baryonrate}
\end{equation}
where $N_C=3$ is the number of collors and $\Gamma_{sph}/V $
is the sphaleron rate per unit volume.
Note that I can easily obtain by integrating
Eqs.~(\ref{eq:fluid equation ii}) and
(\ref{eq:integral of fluid equation i})
that $\int \hat \mu=\int_{wall} \hat\mu+\int_0^\infty \mu=0$. This does not
mean that baryon production vanishes because
the sphaleron rate on the wall Eq.~(\ref{eq:sphaleron rate broken phase})
is exponentially suppressed, indeed to a first approximation I can neglect
the baryon production on the bubble wall to find that the net baryon to entropy
ratio is
\begin{equation}
\frac {n_B}{s}= {135 \ln 2 \over 2 \pi^2\zeta_3}
{\kappa_{sph} \alpha_w^4 \over g_*} \int dz {m^2 g_A Z \over T^2} V(v_w)
\approx 4
{\kappa_{sph} \alpha_w^4 \over
g_*} \int dz {m^2 g_A Z \over T^2} V(v_w)
\label{eq:baryontoentropy}
\end{equation}
where $s=\frac{2\pi^2}{45} g_* T^3$ is the entropy density,
with $g_*\approx 100$  the effective number of degrees of freedom, and
we have restored the units of temperature.
One should not forget that this result is applicable  only in the limit of
efficient transport, {\it i.e.\/} not too close to the speed of sound, when
$\bar D\sim 0$ and only local baryogenesis is possible.

This result is remarkably simple
- all dependence on the wall
thickness and the diffusion constant drops out.
The dependence on the wall velocity is rather weak; it is given by $V(v_w)$ of
Eq.~(\ref{eq:shifted mu}). For modest velocities
$V(v_w)\sim (1-v_w)^{3/2}/(1-3v_w^3 a_{23}/a_{34})$ so that baryon production
becomes more efficient as the velocity increases; the growth stops when the
condition for efficients transport gets violated: $\bar D\sim v_w L$.
The asymmetry is quite large for the top quarks:  $(m_t/T)^2 \sim 1/4$, so
$n_B/s \sim 10^{- 8} \kappa_{sph} \theta_{CP}$ where
$\theta_{CP}$ characterizes the strength of $CP$ violation.
The $m^2$ dependence in Eq.~(\ref{eq:baryontoentropy}) means that,
at least with standard model-like Yukawa couplings,
the top quark dominates the effect.

The calculation of the classical force effect above uses
the opposite (WKB) approximation to those employed in
quantum mechanical reflection calculations (thin walls
\cite{CKNtransport}, \cite{JPTleptons}).
The classical
force calculation is in some respects `cleaner', because
the production of chiral charge and its diffusion are treated
together.
The classical force affects particles from
all parts of the spectrum, mostly with typical energies $E\sim
T$, and  with no preferential direction, while the quantum mechanical
effect comes mainly from
 particles with a very definite ingoing momentum
perpendicular to the wall: $p_z\approx m_H$ (Higgs mass).
The quantum result falls off exponentially with $L$
as the $WKB$ approximation becomes good.
The quantum result also has a $v_w^{-1}$ dependence
coming from the diffusion time in the medium, which
the classical result loses because the force term is
proportional to  $v_w$.

\section{Acknowledgements}

I wish to thank the Organizing Comittee of the 7th Adriatic Meeting
 for making this wonderful even
possible.  The lecture is based on a collaboration with
Michael Joyce and Neil Turok.
The author is supported by funding from PPARC.


\begin{thebibliography}{99}
\bibitem{GuyMoore}{Guy Moore has recently shown that the time
during which sphalerons are active should be
$H^{-1} d(E_{sph}/t)/dT\sim 1000 H^{-1}$, rather than $H^{-1}$; private
communication.}
\end{thebibliography}
\begin{thebibliography}{99}
\bibitem{footnote2}{
The collision integrals vanish identically
for arbitrary background fields at zero $\delta T$, and $\delta v$
respectively, so they take the form $ (\delta T)
\Gamma(Z_z=0,m^2=0)(1 + o(Z_z) + o(m^2) +...)$
But any rate on the R.H.S. multiplies a perturbation, of
$o(m^2 Z_z)$ so that the linear correction to the rate
contributes at  second order in $Z_z$ and can be ignored.}
\end{thebibliography}
\end{document}